\documentclass[twocolumn,showpacs,preprintnumbers,amsmath,amssymb]{revtex4}
\usepackage{graphicx}% Include figure files
\usepackage{dcolumn}% Align table columns on decimal point
\usepackage{bm}% bold math

\begin{document}

%\preprint{SPIE Proceedings}

\title{Self-organized composites of multiwalled carbon nanotubes and \\ nematic liquid crystal 5CB:\\ Optical singularities and percolation behavior in electrical conductivity}% Force line breaks with \\

\author{V.V Ponevchinsky}
\affiliation{Institute of Physics, NAS of Ukraine, 46 Prospect
Nauki, Kyiv 03650, Ukraine}

\author{A.I. Goncharuk}
%\homepage{http://www.Second.institution.edu/~Charlie.Author}
%\email{Second.Author@institution.edu}
\affiliation{Institute of Biocolloidal Chemistry named after F.
Ovcharenko, NAS of Ukraine, 42 Vernadskii Prosp., Kyiv 03142,
Ukraine}

\author{V.I. Vasil'ev}
\affiliation{Institute of Physics, NAS of Ukraine, 46 Prospect
Nauki, Kyiv 03650, Ukraine}

\author{N.I. Lebovka}
\email{lebovka@gmail.com}
 %\homepage{http://www.Second.institution.edu/~Charlie.Author}
\affiliation{Institute of Biocolloidal Chemistry named after F.
Ovcharenko, NAS of Ukraine, 42 Vernadskii Prosp., Kyiv 03142,
Ukraine}%

\author{M.S. Soskin}%
\email{marat.soskin@gmail.com}

\affiliation{%
Institute of Physics, NAS of Ukraine, 46 Prospect Nauki, Kyiv
03650, Ukraine}%

\date{\today}% It is always \today, today,
             %  but any date may be explicitly specified

\begin{abstract}
This work discusses optical singularities and electrical
conductivity behavior in a thin electrooptical cell filled with
composites including multi-walled carbon nanotubes (MWCNTs) and
nematic liquid crystal (LC). The MWCNTs with high aspect ratio
$L/d\approx 300\div1000$  and nematic LC 5CB
(4-pentyl-4-0-cyanobiphenyl) were used. The composites were
prepared by introduction of MWCNTs ($0.0001\div0.1$ \% wt) into LC
solvent with subsequent sonication. The increase of MWCNT
concentration (between $0.005\div0.05$ \% wt) resulted in
self-organization of MWCNTs and formation of micron-sized
aggregates with fractal boundaries. The visually observed
formation of spanning MWCNT networks near the percolation
threshold at ~0.025 \% wt was accompanied with transition from
non-conductive to conductive state and generation of optical
singularities. The observed effects were explained by the strong
interactions between MWCNTs and LC medium and planar orientation
of 5CB molecules near the lateral surface of MWCNTs. It was
speculated that optical singularities arose as a results of
interaction of an incident laser beam with LC perturbed
interfacial shells covering the MWCNT clusters. Behavior of the
interfacial shell thickness in external electric field and in the
vicinity of the nematic to isotropic transition was discussed.
\end{abstract}

\pacs{02.40.-k, 42.25.Ja, 42.30.Ms, 61.30.-v, 61.46.-w, 73.63.Fg, 73.22.-f}% PACS, the Physics and Astronomy
                             % Classification Scheme.
%\keywords{Suggested keywords}%Use showkeys class option if keyword
                              %display desired
\maketitle

\section{\label{sec:level1}%First-level heading:\protect\\ The line break was forced \lowercase{via} \textbackslash\textbackslash
INTRODUCTION} Nanoscience and nanotechnology are popular slogans
of modern science and technique. Nowadays, the most 'hot points'
are related mainly to development of composites including carbon
nanotubes (CNTs). The typical CNTs have nanometer scale diameter
and very high aspect ratio (length-to-diameter ratio) a 1000.
During the last decade, a great interest was also attracted by the
liquid crystalline (LC) composites doped with CNTs ~\cite{Zakri,
Lagerwall2008,Trushkevych,Zhao,Rahman2009} . These materials
display many unique properties and integrate the rod-like
particles (LC and CNTs) with huge difference in aspect ratios. The
CNT+LC nanocomposites are very attractive as objects for
investigation because the highly anisotropic excluded-volume
interactions may result in unique self-organization and new
unexpected effects.

The introduction of CNTs inside LC media may produce many
unexpected effects ~\cite{Saito2001,Lebovka2008,Lysetski2009},
influence the phase transitions in LC
~\cite{Russell2006,Duran2005} and enhance the alignment of LC
~\cite{Dierking2004}. Enhancement of CNT ordering inside LC media
~\cite{Lynch2002, Dierking2005, Lagerwall2007} and sensible
responses of CNTs+LC composites to the shear, external electric or
magnetic field, were also reported
~\cite{Baik,Wang,Srivastava,Shah,Dierking2008}. It allowed
construction of promising electrically or magnetically steered
switches ~\cite{Dierking2004,Dierking2005}. Nowadays, the interest
to the electro-optical properties of CNT+LC composites is
continuously growing ~\cite{Lebovka2008,Lu}. The CNTs essentially
affect the spatial distribution of charges inside LC cells and, as
a result, change their electro-optic response ~\cite{Lee,Huang}.
Different time scales, associated with reorientation of the LC
texture (the short timescale) and with reorientation of the carbon
nanotubes (the long timescale), were identified in the external
electric fields ~\cite{Dierking2008}. The electrical conductivity
and the dielectric constant of LCs doped with carbon nanotubes
demonstrate extraordinary large changes in electric and magnetic
field driven reorientation experiments
~\cite{Dierking2008,Jayalakshmi}.

However, in spite of these vast investigations, the main mechanism
governing structure formation and optical and physical properties
of LC+CNTs composites remained unclear.  Particularly, the typical
micrometer scale length of isolated CNTs (or their aggregates)
dispersed in LC cells is a subject of classical optics. Contrary,
the nm-scale diameter of CNTs belongs to the area of nanooptics.
The nano- and micro-heterogeneities in CNT+LC composites can
produce optical singularities and drastically chance optical
properties of these systems. The optical singularities exist in
variety of natural systems and appear as excellent tool for
investigation of their basic properties ~\cite{Nye,Soskin2001}.
From other side, self-organization in CNT + LC composites at
different concentrations of CNTs can be easily monitored by means
of different experimental techniques, including singular optics
and polarization microscopy methods and measurements of electrical
conductivity.

This work studies optical and electro-physical properties of
multiwalled carbon nanotubes (MWCNTs) and nematic LC 5CB
composites in the concentration range C between 0.0001 and 0.1 \%
wt of MWCNTs.  The behavior of optical singularities, percolation
threshold from  non-conductive to conductive state and electric
field driven effects,  are also investigated and discussed in
details.

\section{\label{sec:LII}%First-level heading:\protect\\ The line break was forced \lowercase{via} \textbackslash\textbackslash
MATERIALS AND EXPERIMENTAL TECHNIQUES}

\subsection{\label{sec:L2.1} Liquid crystal }
The commercially available nematic LC, 5CB (Merck, Germany) was
used as a LC host matrix. An 5CB molecule consists of a rigid
moiety of linked two benzene rings: C$_5$H$_{11}$-\O-\O -C$\equiv$
N. It has a considerable dipole moment,  $\mu\approx 4.76$ D
concentrated on C$\equiv$N group, polarizability $\alpha\approx$
33.1, positive dielectric anisotropy $\Delta \varepsilon\approx
11.7$, and exists in nematic phase within the temperature range
from 295.5 to 308.5 K ~\cite{diMatteo}. In the nematic and
isotropic phases 5CB molecules form dipole -dipole bound dimers
about 2.3 nm long and 0.5 nm thick ~\cite{Luckhurst}. The deep
integration between CNTs and nematic LC and strong anchoring of LC
molecules to the lateral surface of the CNTs
~\cite{Park2007,vanderSchoot2008} is expected owing to the to
strong ?-stacking hexagon-hexagon interactions between benzene
rings and hexagon cells of carbon lattice ~\cite{Jeon2007}.

\subsection{\label{sec:L2.2} Multiwalled carbon nanotubes}
The MWCNTs were produced from ethylene by CVD method, involving
using of FeAlMo as a catalyst (SpecMash Ltd., Ukraine, Fax: 380 44
5010620) ~\cite{Melezhyk2005}. The MWCNTs were further treated by
alkaline and acidic solutions and washed by distilled water until
reaching of distilled water pH in the filtrate. The residual mass
content of the mineral additives was 0.1\%. The specific surface
area $S$ of the tubes was 190 m$^2$/g.

\begin{figure}[!htbp]
\centering\includegraphics*[width=0.4\textwidth]{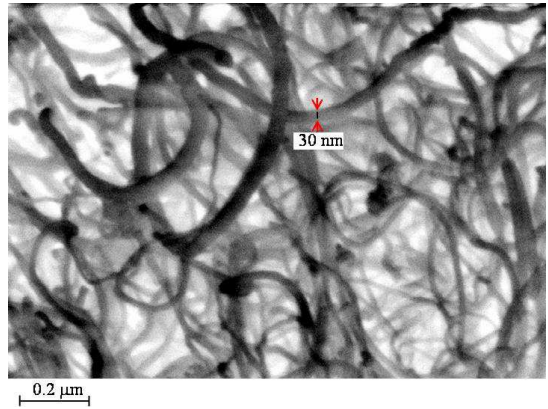}\hfill
\caption {\small Scanning electron microscopy images of
MWCNTs.}\label{fig:L01}
\end{figure}

Fig. 1 shows the scanning electron microscopy (SEM) images of the
MWCNTs in the powder state. High resolution environmental SEM
images were obtained at room temperature using an XL30 ESEM-FEG
instrument (Philips International, Inc., Washington, USA),
operating at the voltage of 15 kV and the pressure of 1.4-3.0
Torr.

Typically, such MWCNTs have a mean outer diameter of $d\approx 20$
nm and a length $l=5\div10 ~\mu$m ~\cite{Melezhyk2005}. The MWCNTs
are presumably metallic-like due to their large diameters. The
specific electric conductivity $\sigma$ of a powder of compressed
MWCNTs (at a pressure 15 of TPa) is about 10 S$^.$cm${-1}$ along
the axis of compression. As we will see, this influences
essentially the optical properties and electroconductivity of
MWCNT+5CB composites.

\subsection{\label{sec:L2.3} Preparation of MWCNT+5CB composites}
The MWCNT+5CB composites were obtained by addition of the relevant
quantities of MWCNTs to liquid 5CB (T=298 K). The weight
concentration of MWCNTs C varied within $0.0001 \div 0.1$ \% wt.
Ultrasonication was used for preparation of homogeneous mixture of
MWNTs. The MWNT suspensions were sonicated for 5 min at a
frequency of 22 kHz and an output power of 150 W using ultrasonic
disperser UZDN-2T. Ultrasonication is an accepted technique for
dispersing the highly entangled or aggregated nanotube samples
~\cite{Cheol2002}, but longer times of high-energy sonication can
introduce defects and decrease the lengths of nanotubes
~\cite{Hilding2003}.

\subsection{\label{sec:L2.4} Sandwich-type LC cells}
The various techniques were previously used for characterization
of the physical properties of a CNT+LC system ~\cite{Trushkevych}.
in a sandwich-like LC cell, including electrical conductivity
~\cite{Lebovka2008}, optical ~\cite{Lysetski2009} and
electrooptical ~\cite{Lysetski2009, Dolgov2008} measurements. In
this work, the sandwich-type cells with different thickness were
used in optical and electrical conductivity measurement
experiments. The thickness of a cell for optical/electrooptical
measurements h was typically 20 and 100 ~$\mu$m to obtain the high
quality optical images in the investigated range of weight
concentration of MWCNTs ($0.0001\div 0.1$ \% wt). Construction of
the used sandwich-type electro-optical LC cell for optical
measurements is shown (Fig.2).

\begin{figure}[!htbp]
\centering\includegraphics*[width=0.4\textwidth]{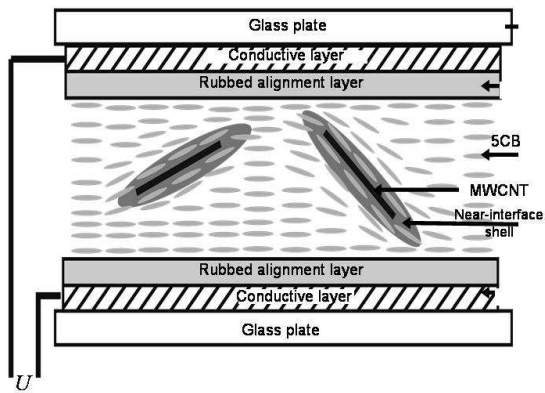}\hfill
\caption {\small Sandwich-type cell for optical/electrooptical
measurements, thickness h is 20 or 100 ~$\mu$m.}\label{fig:L02}
\end{figure}

The interfacial shell, formed by 5CB molecules covering the MWCNT
surface is shown schematically. The conductive TiO$_2$ layers,
covered by polyimide, were used. The thickness of the cells was
set by glass spacers. The polyimide SE150 (Nissan, Japan) layers
were rubbed by a fleecy cloth in order to provide a uniform planar
alignment of LC in the field-out state. The cells were filled by
MWCNT+5CB composite, assembled so that the rubbing directions of
the opposite aligning layers were anti parallel, and then sealed.

\begin{figure}[!htbp]
\centering\includegraphics*[width=0.4\textwidth]{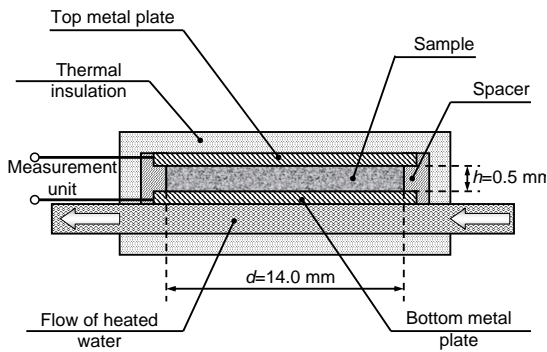}\hfill
\caption {\small Unit for electrical conductivity
measurements.}\label{fig:L03}
\end{figure}

Contrary, large 500 ~$\mu$m gold-covered cells were used for
electrical conductivity measurements (Fig.3). The large thickness
of a cell allows to eliminate direct electrical contacts between
the electrodes through the MWCNT compact bungles and clusters.
Moreover, in the thin cells ($h\simeq 100 ~\mu$m), the size of the
compact clusters is comparable with the size of the cell.
Therefore, electrical conductivity and percolation characteristics
can reflect the crossover $2d-3d$ behavior ~\cite{Lebovka2002}. No
special orienting agents were used in experiments with thick
cells, because they can influence the accuracy of electrical
conductivity measurements. We did not use capillarity method for
filling the cell, because this method may result in size
selectivity of MWCNT clusters and unpredictability of MWCNT
concentration inside the cell.

\subsection{\label{sec:L2.5} Optical microscopy investigations }

The polarization microscope BX51 (Olympus, USA) was used in
optical microscopy investigations. It was equipped by modern
compound microobjectives with micro size focal area and long
enough effective focal length, permitting high-resolution
measurements of optical structure inside a LC cell through 1 mm
thick glass cover sheet. The microscope construction admitted
translation of microscope stage along the optical axis with
precision up to 1 micrometer. This has allowed precise choosing of
the best plane for optical measurements. Moreover, it is useful
for testing the interfacial micro-size shells surrounding MWCNT
aggregates and their 3d structures. The structure of interfacial
shells, surrounding the aggregates, was studied in the quasi
monochromatic regime using the interference filters with 10 nm
FWHM (full width at half maximum). The AC voltage $U$ (up to 11V)
at frequency $f$ of 10 kHz was applied to electrooptical cells
(Fig.2). The 10 kHz AC was used to prevent undesired polarization
phenomena near the electrodes.

\subsection{\label{sec:L2.6} Optical singularities }

Ramified structure of LC interfacial shells is promising for
appearance of optical singularities, born in zero-amplitude points
of a light field full destructive interference ~\cite{Nye,
Soskin2001}. Singularities appear in a natural way in optical
speckle fields with random amplitude and phase distribution
~\cite{Freund2002,Vasil'ev2008}.

It was reasonable to search for them in speckles created in a
laser beam propagating through MWCNT +5CB cell. There are two
types of optical singularities: optical vortices (OVs) in scalar
fields and polarization singularities ~\cite{Nye, Soskin2001}.
Main polarization singularities are C-points.

The arbitrary polarized elliptic light fields possess
simultaneously three kinds of actual OVs:

(i) the Stokes vortices S$_{12}$ located in the intersection point
of Stokes zero lines S$_1$ and S$_2$;

(ii) the opposite circularly polarized ordinary vortex underlying
each C-point; and

(iii) the component vortices of any scalar component of the field.

It was reasonable to search for C-points in the speckle structure
created in a laser beam propagating through 5CB +MWCNT. Their
appearance was checked by the two-arm scheme (Fig.4a).

\begin{figure}[!htbp]
\centering\includegraphics*[width=0.4\textwidth]{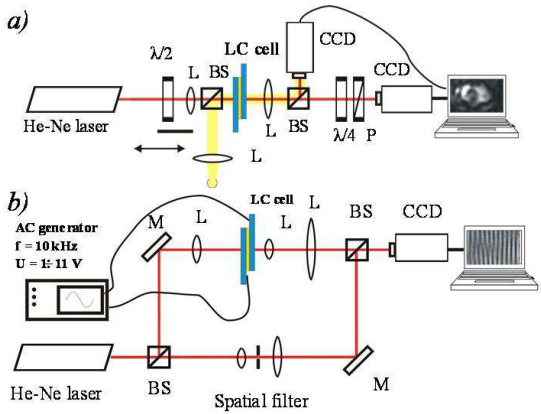}\hfill
\caption {\small The far-field microscopy scheme for detection and
metrology of polarization singularities (a) and optical vortices
(b).}\label{fig:L04}
\end{figure}

The beam was projected on the subsidiary CCD$_1$ camera. A needed
fragment of MWCNT structure was chosen by transverse translation
of the cell. The 5$^x$ microobjective projected the laser beam
transmitted through the cell to the camera CCD$_2$, and Stokes
components were measured.

Optical vortices were detected by the usual two-beam scheme with
plane reference beam (Fig.4b) ~\cite{Soskin2001}. Mirror M$_2$ was
inclined at a small angle to the beam axis for creation of
interference pattern between the reference beam and the object
beam. OVs were detected by well known "fork fringes"
~\cite{Soskin2001}.

\subsection{\label{sec:L2.7} Electrical conductivity measurements }
The electrical conductivity was measured for unaligned samples by
the inductance, capacitance and resistance (LCR) meter 819
(Instek, 12 Hz-100 kHz) in a cell (Fig. 2) equipped with two
horizontal gold electrodes (diameter $d=12$ mm, inter-electrode
space $h = 500~ \mu$m). The applied external voltage was $U =
1.275$ V. The measuring frequency $f$ was chosen within
$1\div10^5$ Hz. In typical experiments, the voltage and frequency
dependences at MWCNT concentrations below the percolation
threshold ($<0.02$ \% wt) (tunneling/hopping regime) and above it
(multiple contacts regime) were investigated.

\subsection{\label{sec:L2.8} Computer assisted image analysis of multiwalled carbon
nanotubes } The binary images were analysed using the box-counting
method, with the help of the image analysis software. The
'capacity' fractal dimension $d_f$ was obtained from dependence of
the number $N$ of boxes necessary to cover the boundary of an
aggregate versus the box size $L$, ~\cite{Feder1988, Smith1996}.
The estimated fractal dimension $d_f$ depicts the morphology of a
checked aggregate in 2d projection. Its value varies between 1
(corresponding to a linear aggregate) and 2 (corresponding to a
compact aggregate). The fractal dimension corresponding to
three-dimensional aggregates $d_{f3}$ can be estimated as $d_{f3}
= d_f + 1$ ~\cite{Feder1988}.

\subsection{\label{sec:L2.9} Statistical analysis}
Each measurement was repeated, at least, five times
to calculate the mean value of the experimental data.

\section{\label{sec:LIII}%First-level heading:\protect\\ The line break was forced \lowercase{via} \textbackslash\textbackslash
RESULTS AND DISCUSSIONS}
\subsection{\label{sec:L3.1} Aggregation of MWCNTs}
Typically, the concentration of MWCNTs $C$ in investigated
composites varied in a wide interval from $10^{-9}$ up to few \%
wt ~\cite{Zakri}. Ultrasonication allowed good dispersion of
MWCNTs to the isolated small bundles.

The strong aggregation tendency of MWCNTs, related to van der
Waals attraction, increases with increase of $C$. The typical
micro photos of MWCNT+5CB composites at different MWCNT
concentrations $C$ between $0.005\div 0.05$ \% wt., along with
data of fractal analysis, are shown in Fig. 5.

\begin{figure}[!htbp]
\centering\includegraphics*[width=0.4\textwidth]{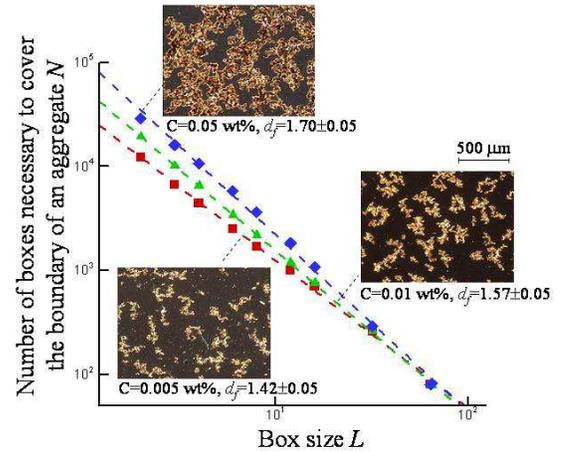}\hfill
\caption {\small Typical micro photos of MWCNT+5CB composites at
different concentrations of MWCNTs and fractal analysis of MWCNT
aggregates using the box-counting method. The data were obtained
in $h =20 ~\mu$m cell, $4^x$, , $T= 297$ K; the polarizer was
oriented along direction of director for planar oriented 5CB
molecules; the analyzer was crossed.}\label{fig:L05}
\end{figure}

At small concentrations of MWCNTs ($C=0.005$ \% wt), the separated
islands of aggregates in a "sea" of nematic 5CB are realized.
These aggregates become thicker and capture sometimes small
nematic "lakes" when weight percentage grows (See pattern for
$C=0.01$\% in Fig.5). The mean aggregate size and distance between
aggregates are of order of $\sim 200 ~\mu$m. They are distributed
randomly in space, which corresponds to the random distribution of
aggregation seeds, formed in initially homogeneous composite.

Taking into account 3d structure of these aggregates and
dimensions of individual MWCNT, $d\approx 20$ nm and $l=5\div10
~\mu$m, each aggregate contains in average up to the million of
nanotubes. At percolation threshold, visually observed near the
concentration $C\approx 0.025$\% the aggregates formed the span
network. At larger concentration of MWCNTs (See pattern for
$C=0.05$ \% wt in Fig.5), the individual aggregates disappeared
and MWCNTs formed dense networks with inner isolated 5CB "lakes".
Borders of aggregates were extremely ramified and fractal analysis
of 2d contours of the aggregates was made (Fig.5). It was
performed on twenty random fragments for each concentration of
nanotubes. The most smooth and most fractal (the highest fractal
dimension $d_f\simeq1.7$) were inner borders of 5CB "lakes" inside
nanotube networks at largest concentration of MWCNTs, $C=0.05$ wt
\%. However, the fractal dimension $d_f$ decreased with decrease
of $C$: $d_f\simeq 1.57$ and $d_f\simeq 1.42$ for aggregates at
concentrations of 0.01 and 0.005 wt \%, respectively.

\begin{figure}[!htbp]
\centering\includegraphics*[width=0.4\textwidth]{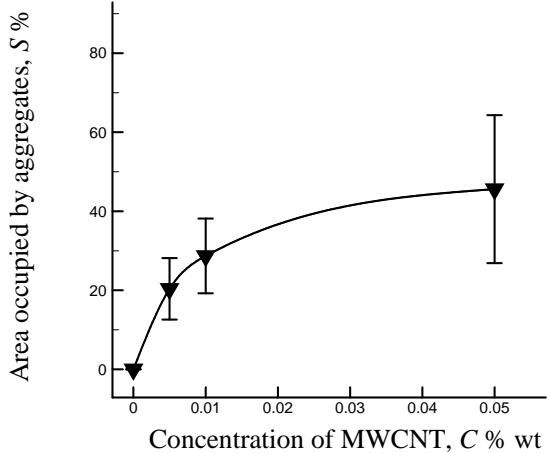}\hfill
\caption {\small Projected area $S$, occupied by MWCNT aggregates,
versus concentration of nanotubes $C$ in MWCNT+5CB composites. The
data were obtained from analysis of micro photos in $h =20 ~\mu$m
cell, $4^x$,, $T= 297$ K; the polarizer was oriented along
direction of director for planar oriented 5CB molecules; the
analyzer was crossed.}\label{fig:L06}
\end{figure}

Fig.6 presents projected area $S$ versus the concentration $C$ of
MWCNTs in $20 ~\mu$m cells, where the solid line corresponds to
the least square fitting of the experimental points (filled
gradients, $\blacktriangledown$) to the power function, i.e.,
$S\propto C^\beta$ , with $\beta\approx 0.34$. Note, that for pure
2d filling, the linear dependence between $S$ and $C$ should be
fulfilled both for fractal and compact objects, i.e.,  $\beta=1$.
For 2d projection of unconfined 3d compact objects, $\beta=2/3$,
while empirical estimates give  $\beta\approx 1$ for fractal
objects; as example, for soot aggregates $d_f=1.82\pm 0.08$ and
$\beta=0.92\pm 0.02$ ~\cite{Koylu1995}. The observed anomalous
behavior with $\beta\approx 0.34<1$, possibly, reflects the
effects of spatial confinements in small 20 ~$\mu$m cells. The
slow growth of $S$ with increasing $C$ ($\beta <1$) reflects
formation of stacked compact aggregates in lateral direction to
the cell wall with small projected area.

\begin{figure}[!htbp]
\centering\includegraphics*[width=0.4\textwidth]{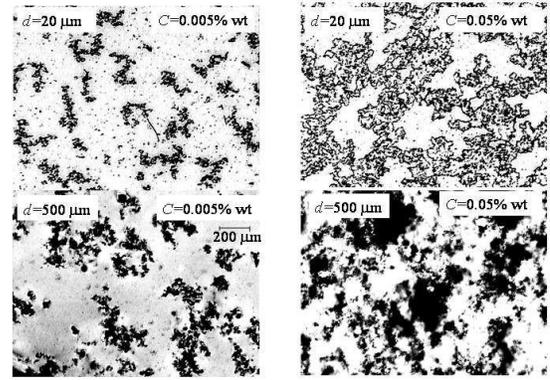}\hfill
\caption {\small Micro photos of aggregates at different
concentrations of MWCNTs in sandwich-like cells with small (h=20
m) and large ($h=500 ~\mu$m) thickness. a - $h= 20  ~\mu$m, $C$ =
0.005 wt\%; b - $h= 20 ~\mu$m, $C = 0.05$ wt\%; c - $h= 500
~\mu$m, $C = 0.005$ wt\%; d - $h= 500  ~\mu$m, $C = 0.05$ wt\%, ,
$T= 297$ K.}\label{fig:L07}
\end{figure}

Fig.7 compares the aggregate structures formed in the cells with
small ($h=20 ~\mu$m) and large ($h=500 ~\mu$m) thickness. Note
that aggregate dimensions can reach multimicrometer size even at
small concentration $C$ of MWCNTs, below the percolation
threshold. For example, at $C=0.005$ \% wt,
 the linear dimensions of aggregates exceed $200 ~\mu$m (Fig.5).
 So, even at small concentration of MWCNTs, the spatial confinements in a small
 thickness cell, $h=20 ~\mu$m, can seriously affect the aggregate structure and bring to
 formation of a stacked compact aggregate. For larger cell with thickness of $h=500 ~\mu$ m, such effects become weaker. Nevertheless, difference in
 a visual perception of the structure of aggregates for
 $C=0.005$ \% wt in both cells was not very significant.
 The important difference in the structure of aggregates,
 formed in the cells with small ($d=20  ~\mu$m) and large ($d=500 ~\mu$m)
 thickness, was observed at large concentration of MWCNTs
 $C=0.05$ \% wt that exceeded the percolation threshold.
 The clusters were visually thicker in a cell with small thickness, $h=20 ~\mu $m.

\subsection{\label{sec:L3.2} Induced polarization singularities and optical vortices}
Fig.8 shows two qualitatively different examples of near- and
far-field structures: mainly clean laser beam core area (Fig.8a)
and pronounced nanotube aggregates (Fig.8c).

\begin{figure}[!htbp]
\centering\includegraphics*[width=0.4\textwidth]{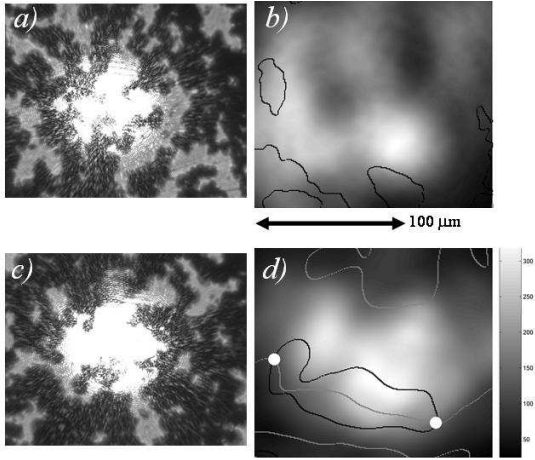}\hfill
\caption {\small Generation optical vortices, induced inside the
laser beam core at different concentrations C of nanotubes in
MWCNT+5CB composites. Negative and positive optical vortices are
marked by white circle and black cross, respectively ($h=100 ~
\mu$m); a - $C =10^{-4}$ wt\%; b - $C =10^{-3}$ wt\%; c - $C
=0.01$ wt\%, d - $C =0.1$ wt\%, $T= 297$ K. }\label{fig:L08}
\end{figure}

As expected, no singularities exist in the clean area (Fig.8b) in
contrary to the pair of C-points appearing on two interception
points of the Stokes S$_1$ and S$_2$ zero-lines on slopes of the
intensity distribution ~\cite{Freund2002} (Fig.8d).

\begin{figure}[!htbp]
\centering\includegraphics*[width=0.4\textwidth]{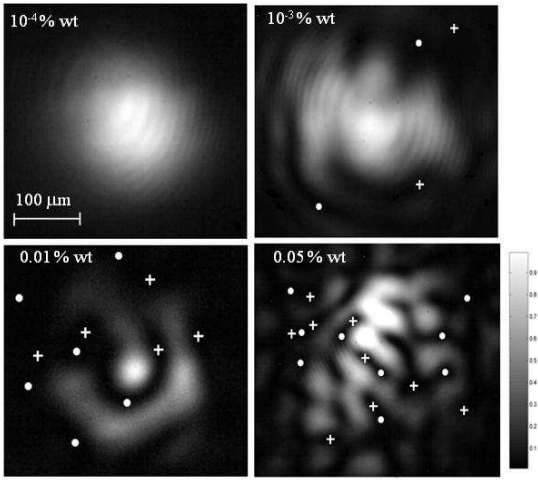}\hfill
\caption {\small The induced circularly polarized singular C
points in MWCNT+5CB composite (0.01 \% wt, $h=20 ~\mu$m, $T= 297$
K). The near-field structure of a laser beam passed through cell
when contours of MWCNT aggregates: are mainly out (a) and within
(c) the beam core. The laser beam core doesn't contain
singularities (b) in contrary, a pair of C points is created (d).
White (dark) grey curves are $S_1$ ($S_2$) zero lines.
}\label{fig:L09}
\end{figure}

The induced optical vortices in MWCNT+5CB cells (Fig.9) at
different concentrations $C$ were checked using the far-field
microscopy scheme (Fig.4b). At very small concentration of MWCNTs
$C<10^{-4}$ wt \%, the optical vortices were absent. They started
to appear at $C\approx 10^{-3}$ wt\% in the periphery of a beam
core (Fig.9b). Further increase of $C$ in the interval between
0.01 and 0.05 wt\% resulted in more speckle structuring of the
beam core and generation of new optical vortices between them
(Fig.9c,d). Enhanced generation of optical vortices in the
vicinity of percolation threshold reflects appearance of spanning
networks. Evidently, such fractal structure of MWCNT aggregates
and highly anisotropic random structure of LC interfacial shells
initiate complicated polarization microstructure of a propagating
light beam. However, direct relation between concentration of
optical vortexes and percolation characteristics is still unclear
and requires more thorough study in future.

\subsection{\label{sec:L3.3} Percolation behavior of electrical conductivity }
The electrical conductivity of MWCNT+5CB composites  $\sigma$ was
within $2-3^.10^{-9}$ S/cm at small concentration of MWCNTs C
(below 0.01 \% wt) and was close to the electrical conductivity of
pure 5CB ( $10^{-9}$ S/cm). With increase of MWCNT concentration C
within 0.01 and 0.1\%, an abrupt growth of by several orders of
magnitude was observed. It evidently reflected the percolation
transition from non-conducting to conducting state at the
percolation threshold concentration $C_c\approx 0.025$ \% wt.

Formation of the percolating structures at $C_c\approx 0.025$ \%
wt was supported also by the observed optical microscopy images.
Namely, at this concentration of nanotubes, the MWCNT aggregates
start to touch and formation of the spanning networks occurs. This
demonstrates close relation between direct optical observations
and behavior of electrical conductivity in MWCNT+5CB composites.

\begin{figure}[!htbp]
\centering\includegraphics*[width=0.4\textwidth]{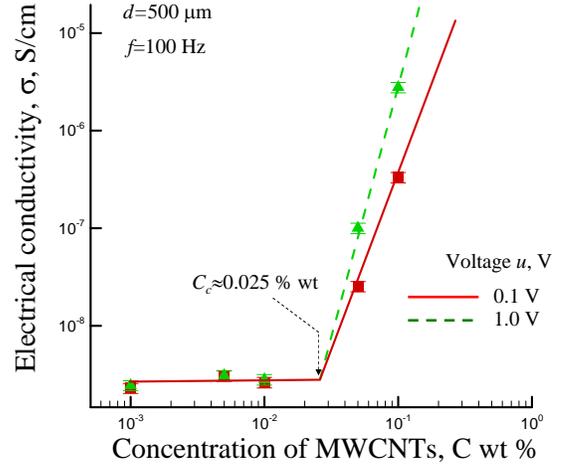}\hfill
\caption {\small The percolation behavior of electrical
conductivity for MWCNT+5CB composites at $C=0.001\div0.1$ \% wt.
$d = 500 ~\mu$m, $f = 100$ Hz, $T= 297$ K. }\label{fig:L10}
\end{figure}

\begin{figure}[!htbp]
\centering\includegraphics*[width=0.4\textwidth]{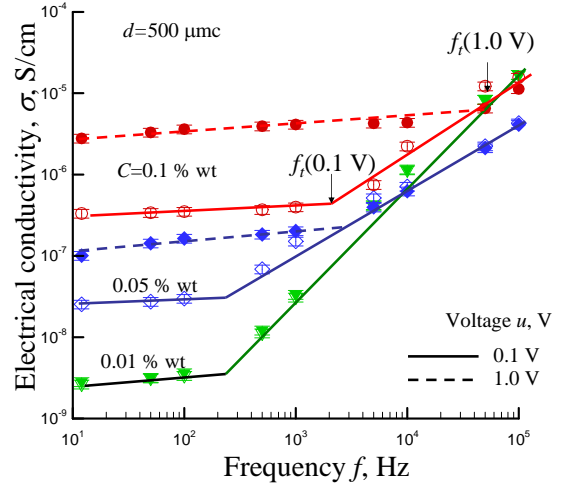}\hfill
\caption {\small Voltage and frequency dependence of electrical
conductivity for 5CB+WCNT composites at $C=0.01\div0.1$ \% wt. $d
= 500 ~\mu$m, $T= 297$ K. }\label{fig:L11}
\end{figure}

The electrical conductivity $\sigma$   in the studied MWCNT+5CB
composites was the increasing function of applied voltage and
frequency (Fig.10). The non-ohmic behavior was also observed in
current-voltage characteristics of MWCNT+EBBA composites
~\cite{Lebovka2008}. It can be explained on the basis of
hopping-tunneling model of transport disruption across the
insulating LC regions between the MWCNTs ~\cite{Mott1971}.

An increase of $\sigma$ with $f$ increase evidenced capacitive
nature of electric conductivity. It is specific for behavior of
the electrical conductivity of a MWCNT composite. Note that the
frequency dependence of $\sigma$ became essential above some
transition frequency value, $f>f_t$, where the scaling law of
$\sigma\propto f^m$ type was observed. The value of $f_t$ was an
increasing function of  MWCNT concentrations $C$. The observed
scaling law of $\sigma\propto f^m$ type evidenced the presence of
the hopping/tunneling mechanism of charge transfer for high
frequencies above $f>f_t$. Above percolation threshold at
$C>C_c\approx 0.025$ \% wt, dependence of $f_t$ vs the applied
voltage $U$ was also observed (Fig.11). For example, $f_t(0.1
V)\approx 200$ Hz and $f_t(1.0 V)\approx 3^.10^3$ Hz at
 $C=0.05$ \% wt and $f_t(0.1 V)\approx 2^.10^3 $ Hz and  $f_t(1.0 V)\approx 4^.10^4 $Hz at
 $C=0.1$  \% wt.
\begin{figure}[!htbp]
\centering\includegraphics*[width=0.4\textwidth]{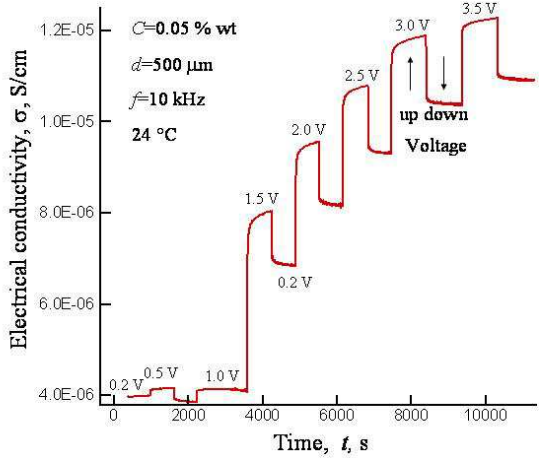}\hfill
\caption {\small Effect of applied voltage on behavior of
electrical conductivity of CNT+5CB composites. $C$=0.05 \% wt,
$d=500 ~\mu$m, $f=10$ kHz, $T= 297$ K.}\label{fig:L12}
\end{figure}

\subsection{\label{sec:L3.4} Electric field driven effects}
The demonstrated dependence of $f_t$ versus the applied voltage
$U$ (Fig.11) evidences the presence of field contribution to the
conductivity mechanism, related with the structure of LC medium
between different MWCNTs. The presence of such electric field
driven effect to the electrical conductivity is supported by the
observed transient behavior of electrical conductivity after
abrupt changing of the applied voltage.

\begin{figure}[!htbp]
\centering\includegraphics*[width=0.4\textwidth]{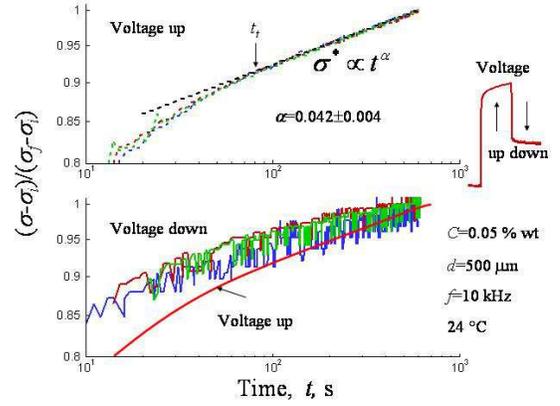}\hfill
\caption {\small Effect of applied voltage abrupt increasing on
behavior of electrical conductivity of MWCNT+5CB composites.
$C$=0.05 \% wt, $d=500 ~\mu$m, $f=10$ kHz, $T= 297$
K.}\label{fig:L13}
\end{figure}

Fig.12 shows time dependence of electrical conductivity in the
experiments with abrupt increase of voltage $U$ from initial value
0.2 V up to 0.5 ($voltage up$ mode) during the time of exposure
$\approx 600\div 1000$ s, followed by it abrupt decrease to the
initial value of 0.2 V ($voltage down$ mode) during the time of
exposure $\approx 600\div 1000$ s and further cyclic voltage
increase up to the final value of 3.5 V in the last up-down
voltage cycle. The electrical conductivity continually increased
and the transient behavior was observed after each such up-down
voltage cycle.

Fig.13 demonstrates such transient behavior of electrical
conductivity in the scaled coordinates
$(\sigma-\sigma_i)/(\sigma_f-\sigma_i)$ vs $t$, where $\sigma_i$
is the initial conductivity (before abrupt change of voltage),
$\sigma_f$ is the final conductivity after exposure time of
$\approx 600$ s, and time t is counted after the abrupt change of
voltage.

It is interesting, that for all the studied processes with jumps
to different values of $U$ during the separate $up-down$ voltage
cycles, the transition curves were falling into the universal
master curves for $voltage ~up$ and $voltage~ down$ modes, and the
curves were smoother for $voltage~ up$ mode than for $voltage~
down$ mode. Two different time processes were evidently present
for the studied system: the faster (at $t<t_t\approx$100 s) and
slower power law process at long exposure times: $\sigma\propto
t^\alpha$ with $\alpha=0.042\pm 0.004$ for $voltage~up$~ mode and
$\alpha=0.03\pm 0.01$ for $voltage~ down$~ mode. The fast process,
possibly, reflects the time relaxation evolving 5CB and MWCNT
coupling, and the slow process may be related with structural
reorganization of MWCNT networks driven by electric field. The
presence of such changes is supported by the observed irreversible
increase of electrical conductivity after the voltage on/off
circles.
\begin{figure}[!htbp]
\centering\includegraphics*[width=0.4\textwidth]{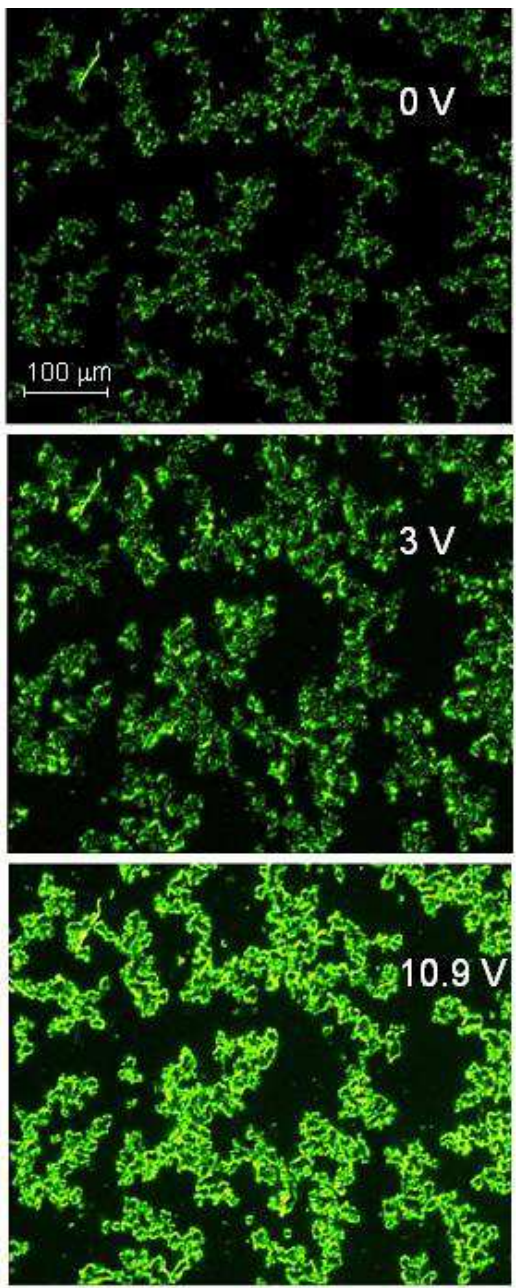}\hfill
\caption {\small Micro photos of 0.05 \% wt MWCNT+5CB composite at
different applied voltages. The data were obtained for $h =20
~\mu$m cell, $4^x$ microobjective; the polarizer was oriented
along direction of director for planar oriented 5CB molecules; the
analyzer was crossed. a - 0 V; b - 5 V; c - 10.9 V, $T= 297$
K.}\label{fig:L14}
\end{figure}

The explanation of the electric field driven transient effect
requires the appropriated physical model. The similar transient
behavior in response of electrical conductivity to on/off
switching of the electric field was previously observed in
suspensions of MWCNTs in distilled water, ethanol, and isopropanol
~\cite{Liu2004}. The observed electric field driven effects can be
related to migration of CNTs and their alignment by electric
fields ~\cite{Yamamoto1998, Chen2001,Nagahara2002,Krupke2003,
Peng2008}. The AC electric charging induces MWCNTs movement in
gradient directions of electric intensity, well known as
dielectrophoresis electrokinetic flow. Consequently, the CNTs can
be rotated and aligned in the LC media by dielectrophoretic force.
This phenomenon occurs, when the complex permittivity of the
suspended CNTs differs from that of the medium.

The strong electric field gradient is also expected near the
surface of ramified MWCNT aggregates. So, dielectrophoresis
mechanism may be appropriate for explanation of the observed
transient phenomena. It was previously shown that MWCNTs were not
only aligned along the field, but they were also migrating
laterally, adding thickness to MWCNT shells, and these processes
were the function of magnitude, frequency, and time of electric
field application ~\cite{Wang2008}.

The strong electric field gradient near the surface of MWCNT
aggregates can noticeably perturb the LC structure in the
interfacial shells surrounding the MWCNT aggregates. This effect
can be directly demonstrated by analyzing the optical microscopy
images of aggregates at different values of applied voltage $U$.

The examples of such images for 0.05 \% MWCNT +5CB composite at
different $U$ are presented in Fig.14. The presence of
enlightenment shells near the surface of MWCNT aggregates clearly
reflected the strong anchoring of 5CB molecules to the surface of
MWCNTs ~\cite{Goncharuk}. The strong anchoring between 5CBs and
lateral surface of MWCNTs is the result of the similarity of
carbon hexagons in 5CB and MWCNTs. The observed enlightenment
shells can be explained by the presence of complicated 3d fields
of elasticity strength inside the layers of anchored 5CBs that
results, in turn, in perturbation of refractive index distribution
near the surface of MWCNTs. The thickness of interfacial shells,
surrounding the MWCNT aggregates, was dramatically dependent upon
the applied voltage U (Fig.15).

\begin{figure}[!htbp]
\centering\includegraphics*[width=0.4\textwidth]{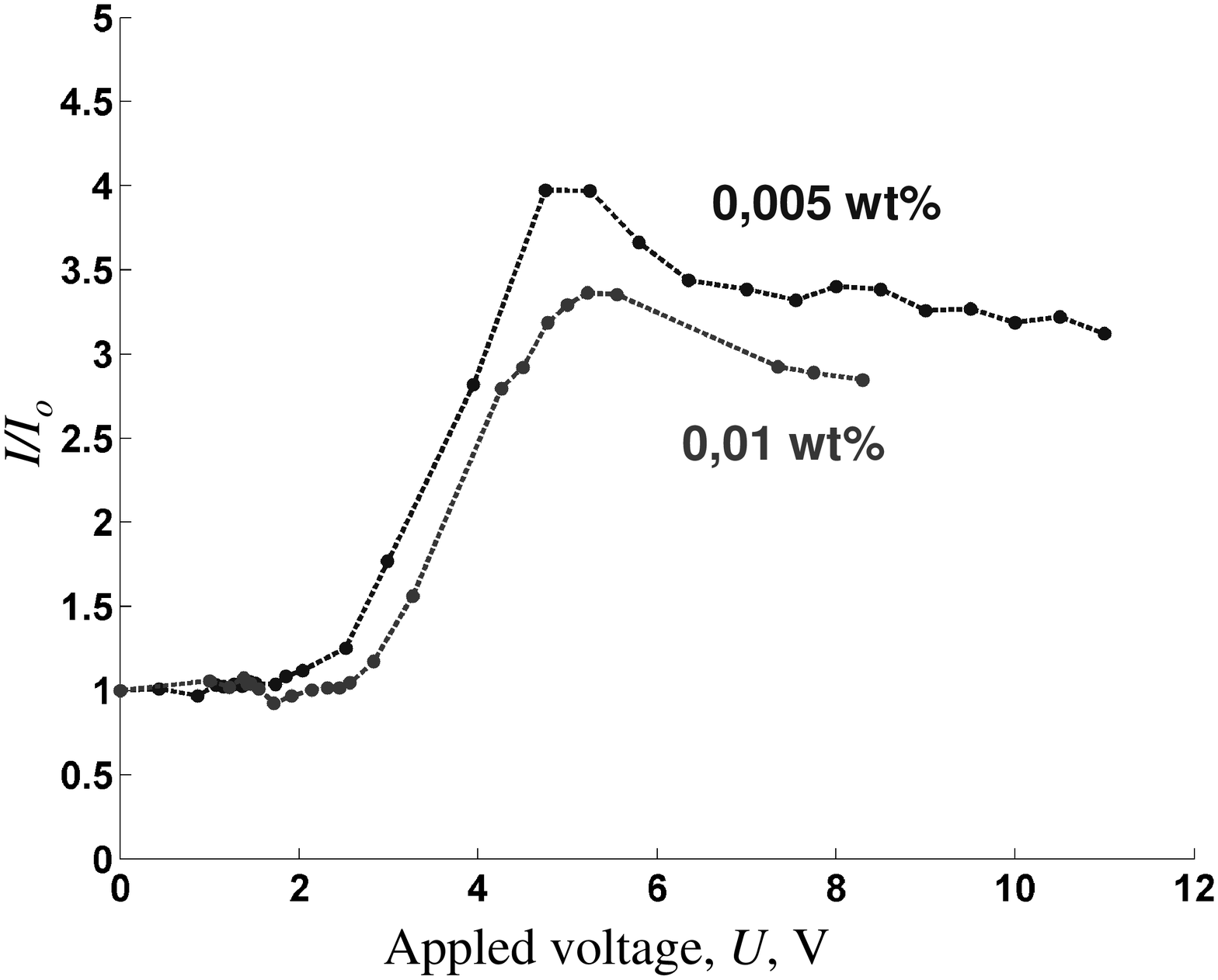}\hfill
\caption {\small Relative integral intensity I/Io versus applied
voltage $U$ for 0.005 \% wt (1) and $C=0.01$ \% wt (2)  MWCNT+5CB
composites. The data obtained for $h =20 ~\mu$m cell, $4^x$, $T=
297$ K; the polarizer was oriented along direction of director for
planar oriented 5CB molecules; the analyzer was
crossed.}\label{fig:L15}
\end{figure}

In order to quantify this effect, the integral image intensity $I$
was calculated by integration of the local image intensity $I_{90}
(x,y)$ over the whole image area: $$I=\int I_{90}(x,y)dxdy$$.
Fig.15 presents dependence of $I/I_0$ for two MWCNT+5CB
composites. The value of $I/I_o$ reached the maximum near some
threshold value of $U\approx 5$ V, and at high voltages it
decreased insignificantly. The qualitatively similar dependences
were observed also for other investigated concentrations of MWCNTs
$C$ in the vicinity of the percolation threshold. The observed
effect of the field on the thickness of interfacial LC shells
surrounding the MWCNT aggregates, possibly, reflects the electric
field driven enhancement of LC structure perturbation in the
interfacial shells and influence of MWCNT loading on the
Frederick's transition inside nematic matrix ~\cite{Blinov}.

\subsection{\label{sec:L3.5} Nematic to isotropic phase transition driven effects}
The phase state of LC matrix may play essential role in
determination of the structure of the interfacial LC shells
surrounding the MWCNT aggregates, their thickness and
electrophysical properties of MWCNT+5CB composites.

\begin{figure}[!htbp]
\centering\includegraphics*[width=0.4\textwidth]{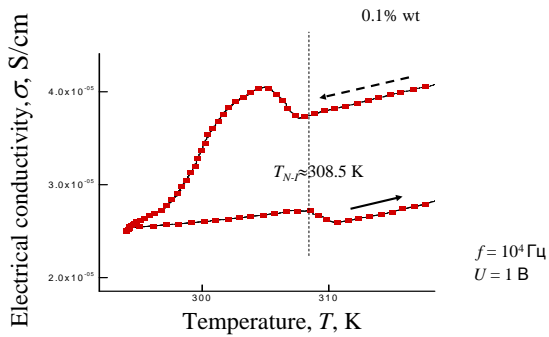}\hfill
\caption {\small Electrical conductivity $\sigma$  versus
temperature $T$ for 0.1 \% wt MWCNT+5CB composites. The data
obtained in $h =500 ~\mu$m cell, $U$=1 V, $f=10^4$ Hz for
un-oriented sample. The arrows show directions of temperature
increase ($\rightarrow$) and decrease
($\leftarrow$).}\label{fig:L16}
\end{figure}

\begin{figure}[!htbp]
\centering\includegraphics*[width=0.4\textwidth]{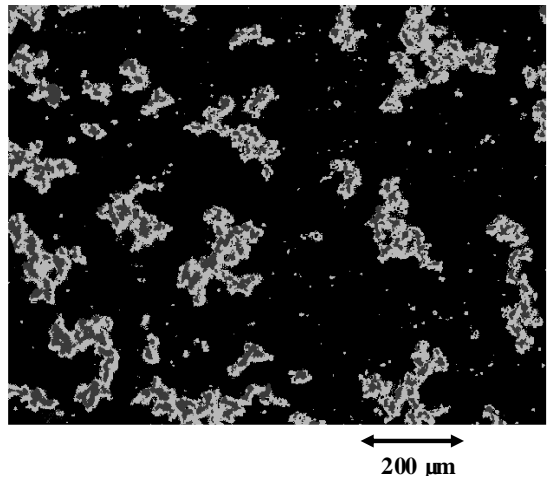}\hfill
\caption {\small Structure of MWCNT aggregates in 5CB nematic
state at 297 °C (white grey) and in the pre-transition state at
305.5°C (dark gray). ($20 ~\mu$m cell, $C=0.01$ \%
wt).}\label{fig:L17}
\end{figure}

Fig. 16 presents the electrical conductivity $\sigma$  versus the
temperature $T$ for 0.1 \% wt MWCNT+5CB composites. The
temperature was initially increased up to 330 K and then decreased
to the room temperature. The electrical conductivity data
demonstrated the presence of distinct non-monotonous behavior in
the vicinity of nematic-isotropic transition ($T_{NI}\simeq
308.5$K) for both thermal heating and cooling regimes. The
electrical conductivity was dropped to lower level at transition
from nematic to isotropic state, and it evidently reflects the
effects of interfacial LC shells, surrounding the MWCNT
aggregates.

It is expected that thickness of interfacial LC shells can
noticeably decrease in the vicinity of phase transition in
isotropic phase, $T\simeq 208.5$ K. Fig. 17 compares the structure
of MWCNT aggregates in nematic state at 297 °C (white grey) and in
the pre-transition state at 305.5°C (dark gray).

Two pronounced changes were shown by comparison of the visual
patterns of aggregates:

(i) "disappearance" of the small islands of MWCNTs, and

(ii) essential shrinkage of area, occupied by the aggregates of
nanotubes (nearly twice shrinkage of their visible dimensions)
with temperature increase.

Both effects can be explained accounting for the decrease of
thickness of the interfacial LC shells, surrounding the MWCNTs. As
a result, single CNTs and their subwavelength bundles are below
resolution of usual optical microscopes.

\section{\label{sec:LV}%First-level heading:\protect\\ The line break was forced \lowercase{via} \textbackslash\textbackslash
CONCLUDING REMARKS }
Up to now, no adequate theories exist for
description of optical and electro-physical properties of MWCNT+LC
composites. The obtained results allow formulation of some
physical model for description of these properties. The essential
property of MWCNT+5CB composites is the presence of a high level
self-organization, formation of MWCNT aggregates and the
interfacial shells (of micrometer thickness), surrounding the
MWCNTs. In particular, in the studied MWCNT+5CB composites, the
strong planar anchoring of 5CB molecules on MWCNT lateral surfaces
was realized by attraction of two benzene carbons to carbon
hexagons of the lateral surface of nanotubes. It induces elastic
strains, including torsion in the neighboring layers of 5CB host,
and produces irregular birefringence, which forms micro size
optical cladding, easily observed through good quality
polarization microscope. Application of the crossed electric field
causes the Frederick's transition of nematic molecules, which
rotate in the plane of incident. The nanotubes tried to align
along E also, which leaded to growth of the cladding dimensions.
The fractal structure of MWCNT aggregates and highly anisotropic
random structure of interfacial LC shells, surrounding the
aggregates, initiates a complicated polarization microstructure of
the propagating light beam and appearance of optical
singularities.

Obtained results can be summarized as follows:

1. The homogenized MWCNT +5CB composites transform in LC cell into
system of volume aggregates of hundreds of thousand nanotubes with
fractal boarders for limited range of MWCNTs content
$0.005\div0.05$ \%wt.

2. 5CB molecules are planar oriented on lateral surface of each
MWCNT. As a result, the macroscopic birefringent irregular
interfacial shell appears. It defines decisively singular and
electrooptical properties of MWCNT +5CB composites. The
interfacial shells surrounding MWCNT disappeared at isotropic
phase what made the small MWCNT clusters invisible in optical
microscope.

3. Laser beam is strongly speckled during the propagation through
heterogeneous birefringent LC interfacial shells and diffraction
on fractal borders of MWCNT aggregates. As a result, polarization
singularities and all types of optical vortices were nucleated.
Electric field influenced strongly the optical singularities
nucleation through the change of LC interfacial shell's
birefringent structure.

4. The studied MWCNT +5CB composites exhibited three known regimes
of electrical conductivity:

(i) tunneling-hopping at C=0.005 %;

(ii) percolation at C=0.025 wt \%; and

(iii) multiple-contacts at C=0.05 wt \%.

The electrical conductivity grew drastically in the vicinity of
percolation threshold, when the gaps between different clusters
disappeared. The electrical conductivity was an increasing
function of the applied voltage U. The frequency dependences of
were also observed for all the C values within 0-0.1 \% wt and
frequency f within 10-105 Hz. This behavior evidences the
hopping/tunneling nature of the mechanism of charge transfer
through LC the interfacial shell, surrounding the MWCNTs.

5. The optical microscopy data evidenced that increase of voltage
U resulted in a drastic increase of thickness  of the interfacial
LC shells, surrounding the MWCNT aggregates, and this effect was
most pronounced at U 5 V in 20  m cell. The presence of electric
field driven effects was also supported by the transient behavior
of electrical conductivity. The two different time processes were
activated by electric field in the voltage on/off circles. The
fast and slow processes presumably reflected the time relaxation
in interfacial shells evolving coupling of MWCNTs and 5CB and the
reorganization of MWCNT network structure, respectively. Moreover,
the thickness of LC interfacial shells, surrounding the
aggregates, was sensitive to phase state of 5CB matrix. This
resulted in a distinct non-monotonous behavior of electrical
conductivity in the vicinity of nematic-isotropic transition (TNI
308.5) and was supported by visually observed essential shrinkage
of area, occupied by interfacial shells and MWCNT aggregates, with
temperature increase.

6. Totality of obtained results started singular optics of CNT+LC
composites.

\section{\label{sec:LVI}%First-level heading:\protect\\ The line break was forced \lowercase{via} \textbackslash\textbackslash
ACKNOWLEDGEMENT} Authors are thankful to Profs. Yuri Reznikov and
Victor Reshetnyak for useful consultations on LC physics, Dr.
Vasil' Nazarenko for polarization microscope Olympus, computer
controlled oven and useful LC discussions, PhD Alex Shumelyuk for
electrical generator and oscillograph. The work was partially
supported by project STCU No. 4687, projects 2.16.1.4, 2.16.1.7
(NAS of Ukraine) and F28.2/058 (FFS, Ministry of Education \&
Science of Ukraine).

%\newpage %Just because of unusual number of tables stacked at end
%\bibliography{apssamp}% Produces the bibliography via BibTeX.

\end{document}